\magnification1200

\rightline{KCL-MTH-04-02}
\rightline{RI 04-01}
\rightline{hep-th/0401196}

\vskip .5cm

\centerline{\bf Kac-Moody Symmetries of Ten-dimensional}
\centerline{\bf Non-maximal Supergravity Theories}

\vskip 1cm

\centerline{ Igor Schnakenburg ${}^2$ and Peter West${}^1$ }

\vskip .5cm

\centerline{\it ${}^1$ Department of Mathematics, King's College
London, UK}

\centerline{\it and}

\centerline{\it ${}^2$ Racah Institute of Physics, The Hebrew
University, Jerusalem, Israel}

\vskip 1cm

\leftline{\sl Abstract}

\noindent
A description of the bosonic sector of ten-dimensional $N=1$ supergravity
as a non-linear realisation is given. We show that if a suitable
extension of this theory were invariant under a Kac-Moody algebra, then
this algebra would have to contain a rank eleven Kac-Moody algebra, that
can  be identified to be
 a particular real form of very-extended $D_8$. We also describe the 
extension of $N=1$ supergravity coupled to an abelian vector gauge field
as a non-linear realisation, and find the Kac-Moody algebra governing the
symmetries of this theory to be very-extended $B_8$. Finally, we discuss
the related points for the $N=1$ supergravity coupled to an arbitrary
number of abelian vector gauge fields.

\vskip .5cm

\vfill

email: igorsc@phys.huji.ac.il, pwest@mth.kcl.ac.uk

\parskip=18pt

\eject


\medskip
{\bf {0. Introduction }}
\medskip

There are no supergravity theories in space-time dimensions greater than 
eleven, but there exists a unique supergravity theory in eleven 
dimensions and two supergravity theories in ten dimensions which differ 
according to whether the two 16 component Majorana-Weyl spinors are of 
same or opposite chirality [1]. The corresponding ten dimensional 
supergravity theories are called IIA [2] and IIB [3] respectively and form 
the low energy effective actions of the string theories with the same name. 
There exists  one other supergravity theory  in ten dimensions which 
possesses a supersymmetry with a Majorana-Weyl spinor that can be coupled 
to Yang-Mills supermultiplets [4]. The quantum version of this theory is  
plagued with anomalies, but these can be canceled by introducing 
Yang-Mills multiplets whose gauge groups are $SO(32)$ or 
$E_8\otimes E_8$ [5]. These two anomaly-free theories are the low energy 
effective actions for the $SO(32)$ and heterotic superstring theories [6].
\par
It is a consequence of supersymmetry that the scalars in supergravity
multiplets belong to non-linear realisations; the first such example was
discovered in  [7]. One of the most celebrated examples concerns  the
four dimensional maximal supergravity where the scalars belong to a
non-linear realisation of $E_7$ [8]. In [9] the coset construction was 
extended to include the gauge fields of supergravity theories. This 
method used generators that are inert under Lorentz transformations and, 
as such, it is difficult to extend this method to include both gravity 
and fermions. The eleven dimensional supergravity theory does not possess 
any scalars and it was widely believed that the symmetry algebras, such as 
$E_7$, found on dimensional reduction were not present in this theory.  
However, it was found that the eleven dimensional supergravity theory 
could be formulated as a non-linear realisation [10]. The infinite 
dimensional algebra involved in this construction was  the closure of a 
finite dimensional algebra, denoted $G_{11}$, with the conformal algebra
in eleven dimensions. The non-linear realisation was carried out by 
ensuring that the equations of motion were invariant under both finite 
dimensional algebras, taking into account that some of their generators 
were in common. The algebra $G_{11}$ involved the space-time translations 
together with an algebra $\hat G_{11}$ which contained $A_{10}$ and the 
Borel subalgebra of $E_7$ as subalgebras. The algebra  $\hat G_{11}$  was 
not a Kac-Moody algebra, however,  it was conjectured [11] that the theory 
could be extended so that the algebra $\hat G_{11}$  was promoted to  
a Kac-Moody algebra.  It was shown that this symmetry would have to contain 
a certain rank eleven Kac-Moody algebra denoted $E_{11}$ [11]. 
\par
The bosonic sector of IIB supergravity theory can also be described as a 
non-linear realisation [12] and the corresponding Kac-Moody algebra is 
$E_{11}$. The massive extension of IIA was formulated as a non-linear 
realisation in [13].
\par
In this paper we construct $N=1$ pure supergravity as a non-linear 
realisation, and identify the underlying Kac-Moody algebra to be $D_8$. 
We then extend the  construction to the case  when the  theory  includes 
a number of  abelian vector field  and find  underlying  the Kac-Moody 
algebra. Finally, we discuss how one might extend this result to the case 
when the vector fields are part of a non-abelian gauge symmetry.


%


%


\medskip

{\bf 1. Pure $N=1$ supergravity}

\medskip

The Lagrangian description of the bosonic sector of $N=1$ supergravity
[4] contains the vielbein $e_a{}^\mu$, the dilaton $A$, and a two form
field $A_{a_1a_2}$. Introducing duals of both of the latter gauge
fields, but not for the gravity, the complete field contents is
given by
$$
h_a{}^b,\quad A,\quad A_{a_1a_2}, \quad A_{a_1\ldots a_6},\quad
   A_{a_1\ldots a_8}.
\eqno(1.1)
$$
We want to identify these fields as the Goldstone bosons of a
non-linear realisation. Thus the symmetry algebra must
contain generators corresponding to the fields in (1.1),
namely
$$
K^a{}_b,\quad R,\quad R^{a_1a_2}, \quad R^{a_1\ldots a_6},\quad
   R^{a_1\ldots a_8}.
\eqno(1.2)
$$
We include the momentum generator $P_a$ in order to introduce the notion
of space-time into the algebra. We note that the $K^a{}_b$ generate
the algebra  $A_9$ which  contains the Lorentz algebra with the
generators
$J_{ab}= K_{ab}- K_{ba}$, where we have raised and lowered indices
with the flat Minkowski metric.
\par
We take the generators introduced above to fulfill the following
commutation relations
$$
  [K^a{}_b,\, K^c{}_d] = \delta^c_b K^a{}_d - \delta^a_d K^c{}_b,\quad
  [K^a{}_b,\, P_c] = -\delta^a_c P_b,
$$
$$
   [K^a{}_b,\, R^{a_1a_2}] = \delta^{a_1}_b R^{aa_2} + \delta^{a_2}_b
   R^{a_1a},\quad  [K^a{}_b,\, R^{a_1\ldots a_6}] = \delta^{a_1}_b
   R^{aa_2\ldots a_6} + \ldots ,
$$
$$
   [R,\, R^{a_1\ldots a_p}] = c_p R^{a_1\ldots a_p},\quad
   [R^{a_1a_2},\, R^{b_1\ldots b_6}] = c_{2,6} R^{a_1a_2b_1\ldots b_6},
\eqno(1.3)
$$
where $\ldots$ stands for the appropriate anti-symmetrisation. We fix
the three constants $c$ to the values
$$
   c_2 = -c_6 = {1 \over 2} = c_{2,6}.
\eqno(1.4)
$$
All not-mentioned commutators (for example $[K^a{}_b,\,R]$) vanish. It
is easy to check that these commutation relations satisfy the Jacobi
identities. We call this algebra $G_{Ips}$.
\par
We take the Lorentz subalgebra to be a local symmetry  while
the whole algebra acts as a  rigid symmetry. More specifically, if we take
the general group element to be
$$
   g = \exp (x^\mu\, P_\mu)\, \exp(h_a{}^b\, K^a{}_b)\,g_A \equiv g_x\,
   g_h\, g_A,
\eqno(1.5)
$$
where
$$
   g_A = \exp({1\over 8!} A_{a_1\ldots a_8} R^{a_1\ldots a_8})\,
   \exp({1\over 6!} A_{a_1\ldots a_6} R^{a_1\ldots a_6})\,
   \exp({1\over2} A_{a_1a_2} R^{a_1a_2})\, \exp(AR),
\eqno(1.6)$$
then we demand that the theory be invariant under
$$
   g\to g_0\, g\, h^{-1},
\eqno(1.7)
$$
where $g_0$ is an element of the whole group $G_{Ips}$ and $h$ an
element of the local Lorentz transformations. This last equation means
that we can always choose a particular Lorentz transformation to
parametrise the resulting group element in terms of Goldstone fields
alone. In turn the Maurer-Cartan form will always transform covariantly
under the local Lorentz group.
\par
We now calculate the Maurer-Cartan forms in the presence of the Lorentz
connection $\omega = {1 \over 2} dx^\mu\omega_{\mu\,b}{}^aJ^b{}_a$:
$$
   {\cal V}= g^{-1}dg -\omega.
\eqno(1.8)
$$
Since the Lorentz connection transforms as
$$
   \omega\to h\omega h^{-1} + hdh^{-1},
\eqno(1.9)
$$
we find that
$$
  {\cal V} \to h{\cal V} h^{-1}.
\eqno(1.10)
$$
Defining
$$
  {\cal V}\equiv dx^\mu(e_\mu{}^a P_a + \Omega_{\mu\,a}{}^b K^a{}_b) +
  dx^\mu (\sum_{p=0,2,6,8}{1\over p!}e^{-c_pA}\tilde D_\mu A_{a_1\ldots
  a_p}R^{a_1\ldots a_p} ) .
\eqno(1.11)
$$
 one finds that
$$
   e_\mu{}^a = (e^h)_\mu{}^a,\quad \Omega_{\mu\,a}{}^b =
   (e^{-1}\partial_\mu e)_a{}^b - \omega_{\mu\,a}{}^b.
\eqno(1.12)
$$
The expression for   $\tilde D_\mu A_{a_1\ldots a_p}$ will be given
below.
\par
Pure $N=1$ supergravity is the non-linear realisation of the group
that is the closure of the algebra $G_{Ips}$ and the conformal algebra
in ten dimensions. We therefore take only those combinations of the
Cartan forms of $G_{Ips}$ that can be rewritten as Cartan forms of the
conformal group. This procedure was explained in reference [10]
where it was shown that there was then a  unique
constraint on  $ \Omega_{ a b}{}^c$ given by 
$$
   \Omega_{a[bc]} - \Omega_{b( ac)} + \Omega_{c(ab)}=0.
\eqno(1.13)
$$
In fact, it results in the usual expression of the spin connection in
terms of the vielbeins of the theory. The $G_{Ips}$ covariant derivative
of the gauge field $A_{a_1a_2}$, for example, in equation (1.11)  given by
$$
   \tilde D_\mu A_{a_1a_2} = \partial_\mu A_{a_1a_2} +
   (e^{-1}\partial_\mu)_{a_1}{}^b A_{ba_2} +
   (e^{-1}\partial_\mu)_{a_2}{}^b A_{a_1b}
\eqno(1.14)
$$
is neither totally anti-symmetrised nor does it contain the
derivatives that occur in general relativity. Demanding that the
theory is invariant under the conformal group and $G_{Ips}$ given in
(1.3,4), we find that we must only use the totally anti-symmetrised
Cartan forms of the gauge fields of equation (1.11). They are
$$
   \tilde F_a = \tilde D_a A = \partial_a A,\ \   \tilde F_{a_1a_2a_3} =
3 e^{-1/2 A} \tilde D_{[a_1}A_{a_2a_3]}
\eqno(1.15)
$$
for the scalar and  the 2-form potential respectively. Similarly we have
$$
   \tilde F_{a_1\ldots a_7} = 7 e^{1/2 A} \tilde
   D_{[a_1}A_{a_2\ldots a_7]}
\eqno(1.16)
$$
for the 6-form potential and
$$
   \tilde F_{a_1\ldots a_9} = 9 (\tilde D_{[a_1}A_{a_2\ldots a_9]} -
   2\cdot 7 A_{[a_1a_2}\tilde D_{a_3}A_{a_4\ldots a_9]} )
\eqno(1.17)
$$
for the 8-form potential, where $\tilde D_a A_{a_1\ldots a_p} =
e_a{}^\mu (\partial_\mu A_{a_1\ldots a_p} + (e^{-1}\partial_\mu
e)_{a_1}{}^b A_{ba_2\ldots a_p} +\ldots )$ is the $G_{Ips}$ covariant
derivative, and $\ldots$ indicates the action of $(e^{-1}\partial_\mu
e)$ on all the other indices of the relevant gauge field. The
equations of motion for the gauge fields can only be built from the
covariant Cartan forms we have found and, assuming  that they must
result in equations that are of the usual order in space-time
derivatives, they  have to be
$$
   \tilde F^{a_1a_2a_3} = {1\over 7!} \epsilon^{a_1\ldots a_{10}}
   \tilde F_{a_4\ldots a_{10}} ,\quad {\rm and}\quad \tilde
   F^{a_1} = {1\over 9!} \epsilon^{a_1\ldots a_{10}} \tilde
   F_{a_2\ldots a_{10}}.
\eqno(1.18)
$$
The only other non-trivial equation of motion involves the covariant
derivatives of the spin-connection, namely the Riemann tensor
$$
   R_{\mu\nu a}{}^b\equiv \partial_\mu \omega_{\nu a}{}^b +
   \omega_{\mu a}{}^c\omega_{\nu c}{}^b - (\mu\leftrightarrow \nu)
\eqno(1.19)
$$
or its contractions and is given by
$$
   R_{\mu\nu}- {1\over 2}g_{\mu\nu}R - {1\over 2} \partial_{(\mu} A
   \partial_{\nu)} A + {1\over 4}g_{\mu\nu} (\partial A)^2
$$
$$
   +{1\over 16} e^{-A}\tilde F_{(\mu}{}^{\mu_1\mu_2}
   \tilde F_{\nu)\mu_1\mu_2} - {1\over 96} e^{-A} g_{\mu\nu}
   \tilde F^{\mu_1\mu_2\mu_3} \tilde F_{\mu_1\mu_2\mu_3} = 0.
\eqno(1.20)
$$
The combinatorial factors in front of each term in the last equation
are not determined by the calculation above and we have simply
inserted the correct values which are a consequence of
supersymmetry. However, it is likely that the theory which is
invariant under the full Kac-Moody algebra discussed later does have
these coefficients fixed by the symmetry. The above are indeed the
equations of motion of pure $N=1$ supergravity [4].
\par
Our next task will be to embed the algebra $G_{Ips}$ (where $Ips$
stands for $N=1$ pure supergravity) into a Kac-Moody algebra which, with
suitable modifications to the theory, is expected to be a symmetry. 
 The procedure was set out in reference [11] and 
as in this reference  we exclude the translation $P_a$ from this
construction.. Let us
therefore split the generators into
$$
   G^+_{Ips} = \{ K^a{}_b,\, a< b,\, a, b =1,\ldots, 10,\,\,\,
   R^{a_1a_2},\,\,\, R^{a_1\ldots a_6} \}
\eqno(1.21)
$$
and the commuting generators
$$
   G^0_{Ips} = \{ H_a = K^a{}_a - K^{a+1}{}_{a+1},\, a = 1,\ldots, 9,
   \,\, D = \sum_a K^a{}_a,\,\,\, R\}
\eqno(1.22)
$$
plus the remaining generators $K^a{}_b$, where $a>b$. The generators
$K^a{}_b$ for $a<b$ are the positive roots of the $SL(10)$ algebra, the
generators $H_a$ are its Cartan generators and the  $K^a{}_b$,
$a>b$  the negative roots. The Lorentz generators are the difference
between positive and negative roots of $SL(10)$. It was proposed in
reference [11] that the local subgroup is that left invariant under the
Cartan involution. This involution transforms positive roots ($E_a$) into
minus the negative roots ($-F_a$) and negative roots ($F_a$) into minus
the positive ones ($-E_a$), apart from $H_a \to - H_a$. The Lorentz
generators are the subset of the $SL(10)$-generators which is invariant
under the Cartan involution. In the bigger algebra $G_{Ips}$ there will
be other invariant generators thus contributing toward the local
subgroup.
\par
Kac-Moody algebras are defined in terms of their simple roots and
their Cartan subalgebra. The positive (negative) roots are then
given by taking repeated commutators of simple positive (negative)
roots subject to the Serre relations. With the above choice of local
subgroup, the positive roots of the Kac-Moody algebra must correspond
to Goldstone fields in the non-linear realisation and as such the
generators in (1.21) must be positive roots of the underlying
Kac-Moody algebra. The set of generators in equation (1.21) can be
found by taking multiple commutators of
$$
   E_a = K^a{}_{a+1},\quad {\rm for}\quad a = 1,\ldots,9,\quad {\rm
   and} \quad E_{10}= R^{9\,10},\quad E_{-1}= R^{5\,6\,7\,8\,9\,10}.
\eqno(1.23)
$$
We can identify these as simple roots of the Kac-Moody algebra. We also
identify the generators of equation (1.22) as the Cartan subalgebra of
the Kac-Moody algebra. We note that the number of simple roots is the
same as the number of Cartan elements, namely eleven.
\par
As explained in reference [11] to identify the Kac-Moody algebra
requires further thought, as it does not contain most of the negative
simple roots. To proceed further we consider certain subalgebras of
$G_{Ips}$.  One such subalgebra is of course $A_9$ corresponding to the
$K^a{}_b$ generators of $SL(10)$. Other simple subalgebras occur when
restricting  the range of the indices. One can easily check that the
algebra (1.3)  contains the Borel subalgebra of $D_6\times A_1$, when
the indices only  assume values $i,j = 5,6,7,8,9,10$. The $D_6$ Borel
subalgebra is  directly realised by the generators $K^i{}_j$ and
$R^{ij}$ and the  6-form generator which in six dimensions can be
dualised to become a scalar and is then identified as the positive
simple root of $A_1$. The seven Cartan generators needed to produce
the Cartan matrix of $D_6\times A_1$ are given uniquely and occur
naturally. They will be given  below.
\par
As just seen, the elements of $G_{Ips}$ contain the algebra $A_9$ and
the (Borel subalgebra of) $D_6\times A_1$. Since the $K^i{}_j$
generators of the latter coincide with part of the generators of the
former (the $A_5$ subalgebra in $D_6$), it is simple to just extend
the $A_5$ subalgebra to the full $A_9$. The additional four Dynkin
nodes correspond to the indices on $K^i{}_j$ also assuming values $i,j
=1,2,3,4$. It is less clear, how to connect the remaining $A_1$ node
of the subalgebra $D_6 \times A_1$ to the resulting Dynkin diagram. We
know, however, that in all known cases (including IIA and IIB
supergravity in ten dimensions, eleven dimensional supergravity, and
the 26 dimensional effective bosonic string action), supergravities
dimensional restriction to two dimensions (indices assuming values
$3,\ldots,d$) affinises the symmetry algebra. Assuming a similar
mechanism to hold for the present case the transition from including
seven dimensions to include eight dimensions then has to produce
affine $D_8$. This fixes the additional $A_1$ node uniquely to
be attached to the sixth node from the right. The whole symmetry
algebra is then found to be very-extended $D_8$. A more complete
argument for the occurrence of very-extended $D_8$ will be given
shortly.
\par
Before identifying other subgroups of $G_{Ips}$, the decomposition of
$SO(n,n)$ into representations of $SL(n)$ is given. The reason for this
will become apparent in due course. The $n(2n-1)$ generators of the
former algebra decompose into generators $K^a{}_b$, $D$, $R^{ab}$, and
$R_{ab}$. The corresponding degrees of freedom add up as
$$
    n(2n-1) = (n^2 - 1) + 1 + {n(n-1)\over 2} + {n(n-1)\over 2}
\eqno(1.24)
$$
respectively. The $K^a{}_b$ generators belong to the adjoint
representation of $SL(n)$ while $D$ is the trace part of $GL(n)$.
The two indexed generators belong to the two-index antisymmetric
representation of $SL(n)$. The generators $R^{ab}$ and $R_{ab}$
transform as 2-tensors under the $A_{n-1} \sim SL(n)$ subalgebra. They
can also be identified as part of the positive and negative roots of
$SO(n,n)$ respectively.
\par
After this comment, we allow the indices of the algebra $G_{Ips}$ in
(1.3) to take values $i,j = 4,\ldots, 10$. We introduce the slightly
redefined generators
$$
   \hat K^i{}_j = K^i{}_j - {1\over 8}\, \hat D + {1\over 2}\, R,
\eqno(1.25)
$$
where $\hat D= \sum_{l=4}^{10}K^l{}_l$. We can dualise the
six-form generator and introduce it into an extra generator by
defining
$$
   \hat K^{-1}{}_i= S_i= {1\over 6!} \epsilon_{ii_1\ldots i_6}
   R^{i_1\ldots i_6}.
\eqno(1.26)
$$
In order to enhance the Borel subalgebra of $SL(7,{\bf R})$ to the
Borel subalgebra of $SL(8,{\bf R})$ we also need to define a new
Cartan element:
$$
   \hat K^{-1}{}_{-1} = {1\over 8}\hat D - {1\over 2} R.
\eqno(1.27)
$$
The index $-1$ corresponds to the new node in the Dynkin diagram
required by the enhancement. Actually, we realise that the combination
${1\over 8}\hat D - {1\over 2} R$ is precisely the one that turns up
in equation (1.25), so we can rewrite the definition of the hatted
$\hat K$'s as
$$
   \hat K^i{}_j = K^i{}_j - \hat K^{-1}{}_{-1}.
\eqno(1.28)
$$
These generators fulfill
$$
   [\hat K^i{}_j,\, \hat K^l{}_m] = \delta_j^l \,\hat K^i{}_m
   -\delta_m^i \,\hat K^l{}_j \quad {\rm for}\quad i,j,l,m =
   -1,4,\ldots,10.
\eqno(1.29)
$$
The negative roots $K^i{}_{-1}$ are missing, but they are not part of
the Borel sub-algebra we seek. We recognise (1.30) to  be the
commutation relations of $SL(8,{\bf R})$. However, we also have the
anti-symmetric two-form generator $R^{ij}$ fulfilling the relations
$$
   [\hat K^i{}_j,\, R^{lm}] = \delta_j^l\, R^{im} + \delta_j^m R^{li},
$$
$$
   [R^{ij},\, R^{lm}] = 0.
\eqno(1.30)
$$
These relations together with (1.29) form the correct commutation
relations of the Borel subalgebra of $SO(8,8)$ when written with
respect to its $SL(8,\,{\bf R})$ subgroup provided we can achieve
the index enhancement in the $R^{lm}$ generators to also include
the $-1$ index. In this case,  we will have shown that the Borel
subalgebra of $SO(8,8)$ is a symmetry of the theory. At first glance
the supergravity theory does not seem to have any further fields
though. It has been discussed  in references [11,12] that for the
theory to be invariant under a Kac-Moody algebra requires it to be
formulated in a way that  treats the gravity and gauge field sectors
of the theory on an equal footing. This requires a formulation of
gravity that mirrors the dual formulation employed for the gauge
fields. As explained in reference [11] one must introduce in D
dimensions the field $A_{a_1\ldots a_{D-3}}{}^b$ along with the usual
field $h_{a_1}{}^b$. Thus in ten dimensions we would have to introduce
$A_{a_4\ldots a_{10}}{}^b$. More evidence for the occurrence of this
field was given in references [11,12] for the cases of eleven dimensional
supergravity, and the two maximal (non-massive) supergravities in ten
dimensions: IIA and IIB. In reference [11] it was shown that there is
indeed a dual formulation of gravity which on the linearised level
involves  the fields $h_{a_1}{}^b$ and  $A_{a_4\ldots a_{10}}{}^b$.
\par
As such, we introduce the corresponding generator
$R^{a_4\ldots a_{10},b}$ which occurs in  the commutation relation
between the six and the two-form potential (1.3) which we modify to
now be
$$
   [ R^{a_1\ldots a_6},\, R^{b_1b_2}] = -c_{2,6} R^{a_1\ldots
   a_6b_1b_2} + {2\over 7} \, R^{a_1\ldots a_6[b_1,b_2]}.
\eqno(1.31)
$$
The coefficient in front of the last generator in this equation can be
chosen by appropriately rescaling this generator and we choose
${2\over 7}$ for later convenience. The change of this commutation
relation is similar to the cases of IIA and IIB supergravity in ten
dimensions.
\par
We now investigate the consequences of this new generator for the
restriction discussed above. In particular, we can dualise this new
generator $T^j = {1\over 7!} \epsilon_{i_4\ldots i_{10}} R^{i_4\ldots
i_{10},j}$ and identify
$$
   R^{-1j} = T^j = {1\over 7!} \epsilon_{i_4\ldots i_{10}}
   R^{i_4\ldots i_{10},j}.
\eqno(1.32)
$$
The commutators of this new generator with the generators $K^i{}_j$,
$K^{-1}{}_j$, and $R^{-1i}$ follow trivially and indeed lead to  the
correct commutation relations for the Borel subalgebra of $SO(8,8)$;
namely
$$
   [ K^{-1}{}_i,\, R^{jk}] =\delta_i^j\,R^{-1k}-\delta_i^k \,R^{-1j}.
\eqno(1.33)
$$
\par
The unique Cartan generators which lead to the Dynkin diagram of
$D_8$ are given by
$$
   H_a = K^a{}_a - K^{a+1}{}_{a+1}\quad {\rm for}\,\, a = 4,\ldots,
   9,
$$
$$
   H_{10} = K^9{}_9 + K^{10}{}_{10} - 2\hat K^{-1}{}_{-1},\quad H_{-1}
   = - K^4{}_4 + 2 \hat K^{-1}{}_{-1}.
\eqno(1.34)
$$
\par
We now show that the required Kac-Moody algebra that contains
$G_{Ips}$ is very-extended $D_8$ by finding the Kac-Moody algebra that
contains two subalgebras $A_9$ and $D_8$ in the required way. Since
we know how six of the nodes of the  $A_9$ subalgebra are contained in
the $D_8$ subalgebra we can attach the three additional nodes in $A_9$
(indices also take values $1,\,2,\,3$) to the $D_8$ Dynkin diagram in
a line so that they also form the $A_9$ algebra in the correct way.
We are then left to find the number of connections of these new nodes
to the two nodes that represent the 2-form and the 6-form potential.
\par
To resolve this point we need to study  the effect of adding a node to
the Dynkin diagram of $A_{r-1}$ to define the  Dynkin diagram of a new
algebra $g$. Deleting this node obviously  leaves us with the algebra
$A_{r-1}$ and so we are in the situation considered in section three of
reference [15]. We may write the root $\alpha_c$ corresponding to the
additional node as $\alpha_c=\sum_i A_{ci}{(\alpha_c,\alpha_c)\over
2}\lambda_i +x$ in the notation of reference [15] which only
considered the case of simply laced algebras. The adjoint
representation of the enlarged algebra $g$  clearly contains the root
$\alpha_c$ and so the $A_{r-1}$ representation with Dynkin indices
$p_j=-{(\alpha_c,\alpha_c)\over 2} A_{cj}$. Indeed, the simple root
$E_c$ corresponding to the additional node is part of this
representation. In particular,  if the additional node attaches with
one line to the $p$th node from the right in the Dynkin diagram of
$A_{r-1}$, the generator corresponding to the simple root  belongs to
the rank $p$ anti-symmetric tensor of representation $A_{r-1}$. If we
consider the notion of level introduced in reference [15], i.e. the
number of times $\alpha_c$ occurs in a root of $g$,  then clearly
$\alpha_c$ is at level one. However, at higher levels one will in
general find additional representations of $A_{r-1}$ in the adjoint
representation of the enlarged algebra $g$. These can be determined at
low levels by considering the constraints discussed in reference [15].
\par
Hence the way a node attaches to the $A_{r-1}$ line is fixed by the
$A_{r-1}$ representation that the corresponding simple root of the
node belongs to. One example is in eleven dimensional supergravity
where the generator corresponding to the 3-form potential attaches to
the third node from the right of the $A_{10}$ line and thus provides
the typical shape of an exceptional Lie algebra.
\par
Returning to  the present case of heterotic supergravity in ten
dimensions, we see that  the simple positive root $E_{10}=R^{9\,10}$
of $G_{Ips}$ in (1.23) belongs to the rank 2 anti-symmetric tensor
representation of the $A_9$ algebra. As a result, this node must
attach only to the second node from the right of the $A_9$ line. As
such, it cannot attach to the three nodes of $A_9$ which are not in
$D_8$.
\par
A very similar argument  holds for the generator corresponding to
the 6-form potential $R^{5678910}$ in equation (1.23) which in
the heterotic case is a simple root since it cannot be built up by
taking commutators of other simple roots. It belongs to the sixth rank
anti-symmetric representation of $A_9$ and as such can only attach to
the sixth node from the right of the $A_9$ line.
\par
As a result we may conclude that  the Kac-Moody algebra we are
searching is uniquely given by very-extended $D_8$ depicted in {\it
Figure A}. Of course this algebra contains an infinite number of
generators and so we expect an infinite number of fields in the
corresponding non-linear realisation. It is hoped that these will lead 
to new propagating degrees of freedom.

\input epsf.tex
\vbox{
\vskip.2cm
\hbox{\hskip3cm\epsfbox{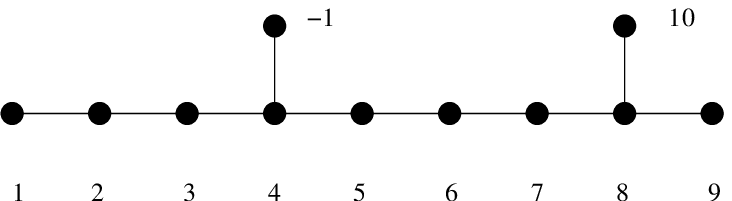}}
\vskip.2cm}
\hskip2cm {\it Figure A: }{The Dynkin diagram of very-extended $D_8$}
\vskip.5cm
\noindent

We find the $A_9$ subgroup as a horizontal line, and the 2-form potential
(the node denoted 10) sticking out from the eighth node, and the 6-form
potential (denoted -1) from the fourth node.
\par
The Cartan generators leading to the usual form of the Cartan matrix of
very-extended $D_8$ are given by
$$
   H_a = K^a{}_a - K^{a+1}{}_{a+1}\quad {\rm for}\,\, a = 1,\ldots,
   9,
$$
$$
   H_{10} = K^9{}_9 + K^{10}{}_{10} - 2 K^{-1}{}_{-1},\quad H_{-1}
   = - K^1{}_1 - K^2{}_2- K^3{}_3 - K^4{}_4 + 2 K^{-1}{}_{-1}.
\eqno(1.35)
$$
The $K^{-1}{}_{-1}$ generator is left unhatted since it contains
the full trace $D= \sum_{a=1}^{10}K^a{}_a$ (compare equation (1.25)). By
construction, the Dynkin diagram of the $D_8$ subalgebra appears very
naturally in {\it Figure A} when considering the indices $-1,4,\ldots,10$
only.
\par
Since $G_{Ips}$ is a symmetry of ten dimensional supergravity it
follows that the Borel subalgebra of $SO(8,8)$ is also a symmetry once
one had succeeded in incorporating the  duals of gravity. The positive
roots $R^{ab}$ of the decomposition (1.24) correspond to Goldstone bosons
of the non-linear realisation. The negative root counterparts $R_{ab}$ do
not have any field analogues and therefore contribute towards the
local symmetry group, and thus should be  hidden symmetries of the
equations of motion.
\par
It is instructive to  check consistency with the similar calculations in
IIA  supergravity theory in 10 dimensions. The field
content of IIA supergravity  including the gauge field duals is given by
$$
   h_a{}^b,\, A_a,\, A_{a_1a_2},\, A_{a_1a_2a_3},\, A_{a_1\ldots
   a_5},\, A_{a_1\ldots a_6},\, A_{a_1\ldots a_7},\, A_{a_1\ldots
   a_8}.
\eqno(1.36)
$$
In pure $N=1$ supergravity, we do not have gauge vectors or three
forms, neither do we have their duals: the seven- or the five form. It
is natural to define an involution ${\cal I}$ on the corresponding
generators which acts as
$$
   {\cal I}: K^a{}_b\to K^a{}_b,\,{\cal I}: R\to R,\,{\cal I}:
   R^{a_1a_2} \to R^{a_1a_2},\, {\cal I}: R^{a_1\ldots a_6},\, {\cal
   I}: R^{a_1\ldots a_8} \to R^{a_1\ldots a_8},
\eqno(1.37)
$$
but also
$$
   {\cal I}: R^a\to -R^a,\,{\cal I}: R^{a_1a_2a_3} \to
    -R^{a_1a_2a_3},\, {\cal I}: R^{a_1\ldots a_5}\to - R^{a_1\ldots
   a_5},\, {\cal I}: R^{a_1\ldots a_7} \to -R^{a_1\ldots a_7}.
\eqno(1.38)
$$
Clearly, ${\cal I}^2 = I$ and the invariant fields are  those
of $N=1$ supergravity. We note that the six-form generator which is
fundamental to $N=1$ supergravity in ten dimensions since it is a
simple positive root, also turns up in IIA supergravity. However, in
the latter theory it is not fundamental but can be build up by taking
repeated commutation relations between the simple positive roots of IIA
which are given by
$$
   K^a{}_{a+1}\quad {\rm for}\,\, a =1,\ldots,9,\quad R^{10},
   \quad {\rm and}\,\,R^{9\,10}.
\eqno(1.39)
$$
Using the $G_{IIA}$ commutation relations
$$
   [R^{a_1\ldots a_p},\, R^{a_{p+1}\ldots a_{p+q}}] = c_{p,q}
   R^{a_1\ldots a_{p+q}}
\eqno(1.40)
$$
with non-trivial coefficients $c_{1,2}$ and $c_{3,3}$ (amongst others)
and the tensor transformation behaviour under $SL(10)$ encoded in
$$
   [K^a{}_b,\, R^{a_1\ldots a_p}] = p\,\delta^{[a_1}_b R^{|a|a_2\ldots
a_p]},
\eqno(1.41)
$$
we find that we can build up the 6-form generator $R^{56789\,10}$ by
taking  repeated commutation relations of the IIA simple positive roots
$$
   2\times R^{10},\, 2\times R^{9\,10}, \, 3\times K^9{}_{10},\,
   4\times K^8{}_9,\, 3\times K^7{}_8,\, 2\times K^6{}_7,\, {\rm
   and}\,\,\, K^5{}_6.
\eqno(1.42)
$$
We do not consider the simple positive root $R^{10}$ in the IIA algebra
since it  is not invariant under the involution $\cal{I}$, but we
introduce a new  simple positive root corresponding to the 6-form
potential which is  invariant under $\cal{I}$. This new simple positive
root $\alpha_{\#}$ can
thus be written as
$$
   \alpha_{\#} = 2\alpha_{10} + 2\alpha_{11} + 3\alpha_9 + 4\alpha_8 +
   3\alpha_7 + 2\alpha_6 +\alpha_5,
\eqno(1.43)
$$
where the subscripts correspond to the following labels in the Dynkin
diagram of $E_8$. The $E_{11}$ of the IIA algebra is just very
extended $E_8$ and the embedding of the root into the larger algebra is
trivial.

\vbox{
\vskip.2cm
\hbox{\hskip3cm\epsfbox{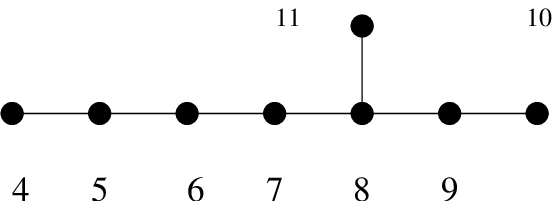}}
\vskip.2cm}
\hskip2cm {\it Figure B: }{The Dynkin diagram of $E_8$}
\vskip.5cm
\noindent
We find that the new root satisfies $\alpha_{\#}^2 =2$, while all the
relations with other roots (but cancelling $\alpha_{10}$, which in IIA
is introduced via $K^a{}_{11} = 2R^a$, and refers to the vector
generator which is not invariant under ${\cal I}$) give the root
system of $D_8$.
\par
We conclude this chapter by noting that the same calculation goes
through for any dimension $d$ of the very-extended $D_{d-2}$ Kac-Moody
algebra. The corresponding gravity theory comprises gravity, as well
as a dilaton and a two-form gauge field along with their duals. One 
example is the case of an effective description of purely bosonic 
string theory in 26 dimensions, which has the same field content, to 
give a more streamlined derivation of the Kac-Moody algebra $K_{27}$, 
or very-extended $D_{24}$ [16].


%


%


\medskip

{\bf 2. $N=1$ supergravity with one vector multiplet}

\medskip

In this section we extend the discussion of the previous section to
the case of  $N=1$ supergravity coupled to one abelian $N=1$ super vector
multiplet. The bosonic field content of this theory is that of the
pure supergravity case, but we have one additional vector field
$A_a$. We denote the corresponding group in the non-linear realisation
by $G_{Ia}$, where the $Ia$ reminds us that we are still dealing with
an $N=1$ theory with an abelian extension. We introduce the Hodge dual
of this vector field, which in ten dimensions can only be a seven-form
field. Introducing generators to all gauge fields, we can write the
general group element of $G_{Ia}$ as
$$
   g = \exp (x^\mu\, P_\mu)\, \exp(h_a{}^b\, K^a{}_b)\,g_A \equiv g_x\,
   g_h\, g_A,
\eqno(2.1)
$$
where $g_A$ is now slightly modified in comparison to (1.5) by the
additional term containing the abelian gauge field
$$
   g_A = \exp({1\over 8!} A_{a_1\ldots a_8} R^{a_1\ldots a_8})\,
   \exp({1\over 7!} A_{a_1\ldots a_7} R^{a_1\ldots a_7})\,
   \exp({1\over 6!} A_{a_1\ldots a_6} R^{a_1\ldots a_6})
$$
$$
  \times\exp({1\over2} A_{a_1a_2} R^{a_1a_2})\,\exp( A_a
  R^a)\,\exp(AR).
\eqno(2.2)
$$
This group element contains all the fields of the theory, but it also
contains the Lorentz group as the antisymmetric part of the $K_{ab}$
generators. As usually, we have added in the momentum generator.
\par\noindent
The generators are taken to fulfill the following commutation relations
$$
  [K^a{}_b,\, K^c{}_d] = \delta^c_b K^a{}_d - \delta^a_d K^c{}_b,\quad
  [K^a{}_b,\, P_c] = -\delta^a_c P_b,
$$
$$
   [K^a{}_b,\, R^c] = \delta^{c}_b R^{a},\quad
   [K^a{}_b,\, R^{a_1\ldots a_p}] = \delta^{a_1}_b R^{aa_2\ldots a_p}
   + \ldots ,
$$
$$
   [R,\, R^{a_1\ldots a_p}] = c_p R^{a_1\ldots a_p},\quad
   [R^{a_1\ldots a_p},\, R^{b_1\ldots b_q}] = c_{p,q} R^{a_1\ldots
   a_pb_1\ldots b_q},
\eqno(2.3)
$$
where $\ldots$ in the second line stands for the appropriate
anti-symmetrisations. Since the case of pure supergravity is a
truncation of the present one, the constants $c_p$ and $c_{p,q}$ should
contain those of pure supergravity (equation (1.4)). We adopt those old
values:
$$
   c_1 = -c_7 = {1\over 4},\quad {\rm and}\quad c_2 = -c_6 = {1\over
   2},
$$
but also:
$$
   c_{1,1} = -2 ,\quad c_{1,6} = 1,\quad c_{1,7} = -{1\over 2},\quad
   {\rm and}\quad c_{2,6} = {1\over 2}.
\eqno(2.4)
$$
Using these commutation relations we can presently calculate the
Maurer-Cartan forms in the presence of the Lorentz connection
(similarly to the previous chapter):
$$
  {\cal V}\equiv dx^\mu(e_\mu{}^a P_a + \Omega_{\mu\,a}{}^b K^a{}_b) +
  dx^\mu (\sum_{p=0,1,2,6,7,8}{1\over p!}e^{-c_pA}\tilde D_\mu
  A_{a_1\ldots a_p}R^{a_1\ldots a_p} ).
\eqno(2.5)
$$
After taking the closure with the conformal group, the objects in
front of the gauge generators are totally anti-symmetrised. They are
given by
$$
   \tilde F_a = \tilde D_a A,
\eqno(2.6)
$$
$$
   \tilde F_{a_1a_2} = 2 e^{-1/4 A} \tilde D_{[a_1} A_{a_2]},
\eqno(2.7)
$$
$$
   \tilde F_{a_1a_2a_3} = 2 e^{-1/2 A} (\tilde D_{[a_1} A_{a_2a_3]} - 2
   A_{[a_1}\tilde D_{a_2} A_{a_3]})
\eqno(2.8)
$$
for the original fields, and for the duals of those fields by
$$
   \tilde F_{a_1\ldots a_7} = 7 e^{1/2 A} \tilde D_{[a_1} A_{a_2\ldots
   a_7]},
\eqno(2.9)
$$
$$
   \tilde F_{a_1\ldots a_8} = 8 e^{1/4 A} (\tilde D_{[a_1} A_{a_2\ldots
   a_8]} - 7A_{[a_1}\tilde D_{a_2} A_{a_3\ldots a_8]})
\eqno(2.10)
$$
and
$$
   \tilde F_{a_1\ldots a_9} = 9 (\tilde D_{[a_1} A_{a_2\ldots
   a_9]} - 2\cdot 7A_{[a_1a_2}\tilde D_{a_3} A_{a_4\ldots a_9]} + 4
   A_{[a_1}\tilde D_{a_2} A_{a_3\ldots a_9]} ).
\eqno(2.11)
$$
These field strengths are the correct field strengths of
$N=1$ supergravity with one abelian vector added in. The equations of
motion can be found by demanding that these field strengths have to be
related to each other in a Lorentz covariant manner. As such, they can
only be
$$
   \tilde F^{a_1\ldots a_p} = {1\over (10-p)!} \epsilon^{a_1\ldots
   a_{10}} \tilde F_{a_{p+1}\ldots a_{10}}, \quad {\rm for}\,\, p
   =1,2,3.
\eqno(2.12)
$$
These are the correct field equations of the $N=1$ supergravity theory
coupled to an abelian vector multiplet. As in the preceding section
it can be shown that also the Einstein equation is given uniquely
(except for some coefficients) after taking the closure with the
conformal group and adopting the inverse Higgs constraint
$\Omega_{a[bc]} - \Omega_{b(ac)} + \Omega_{c(ab)} =0$. The Einstein
equation with fitted coefficients then reads
$$
   R_{\mu\nu}- {1\over 2}g_{\mu\nu}R - {1\over 2} \partial_{(\mu} A
   \partial_{\nu)} A + {1\over 4}g_{\mu\nu} (\partial A)^2 -2
   (F_{\mu}{}^\rho F_{\nu\rho} - {1\over4} G_{\mu\nu}F^2)
$$
$$
   +{1\over 16} e^{-A}\tilde F_{(\mu}{}^{\mu_1\mu_2}
   \tilde F_{\nu)\mu_1\mu_2} - {1\over 96} e^{-A} g_{\mu\nu}
   \tilde F^{\mu_1\mu_2\mu_3} \tilde F_{\mu_1\mu_2\mu_3} = 0.
\eqno(2.13)
$$
We conclude  that the above group $G_{Ia}$ indeed is a symmetry of the
theory.
\par
Assuming the theory is invariant under a Kac-Moody algebra we now wish
to identify it. In order to do so, and after excluding the translations
$P_a$ from this discussion, we divide the generators into 
$$
   G^+_{Ia} = \{ K^a{}_b,\, a<b,\, a,b=1,\ldots, 10,\quad R^a,\,
   R^{a_1a_2},\,R^{a_1\ldots a_6},\, R^{a_1\ldots a_7},\,
   R^{a_1\ldots a_8}\}.
\eqno(2.14)
$$
We identify these as some of the positive roots of the underlying
Kac-Moody algebra. Similarly, the generators of its Cartan subalgebra
are taken to be
$$
   G^0_{Ia} = \{ K^a{}_a,\, a= 1,\ldots,10,\, D= \sum_a K^a{}_a,\quad
   R\}.
\eqno(2.15)
$$
The only generators that are missing form equation (2.3) are the
negative roots of the $K^a{}_b$ algebra, where $a>b$.
\par
Again we can identify  a unique set of simple positive roots by finding
those whose repeated commutators give all those in $ G^+_{Ia} $. These
simple roots are
$$
   E_a = K^a{}_{a+1} \quad {\rm for}\quad a=1,\ldots,9,\quad E_{10} =
   R^{10},\quad E_{-1} = R^{56789\,10}.
\eqno(2.16)
$$
Apart from the obvious $SL(10,\, {\bf R})$ subgroup we can also find
a $B_6\times A_1$ subgroup when we restrict the indices to only assume
six values $i,j = 5,\ldots, 10$. The seven and eight form generators
get projected out in this restriction. Indeed, the Borel subalgebra of
$B_6\times A_1$ is realised in the commutation relations of the field
generators if we slightly rescale the vector generator
$R^i \to {R^i\over \sqrt 2}$. We then find
$$
   [K^i{}_j,\, K^l{}_m] = \delta_l^j\, K^i{}_m - \delta^i_m\, K^l{}_j
$$
$$
   [K^i{}_j,\, R^l] = \delta_j^l\, R^i,\quad [K^i{}_j,\, R^{lm}] =
   \delta^l_j \, R^{im} + \delta^m_j R^{li},
$$
$$
   [R^i,\, R^{lm}] =0,\quad [R^i,\, R^j] = -R^{ij}\quad ({\it here:}\,
   {\rm all\,\,indices \,\, 5,\ldots,10}) .
\eqno(2.17)
$$
As in the case of pure supergravity, the six form generator
$R^{56789\,10}$ is the (simple) positive root of the $A_1$ which
decouples from the $B_6$ algebra part.
\par
In what follows we will need  the  decomposition of $SO(n,n+1)$ into
representations of $SL(n)$. The $n(2n+1)$ generators of the former
split up into generators $K^a{}_b$, $D$, $R^{ab}$ and $R_{ab}$,
but also $R^a$, and $R_a$ (indices running from $1,\ldots,n$). The
corresponding degrees of freedom add up as
$$
   n(2n+1) = (n^2-1) + 1 + 2\times {n(n-1)\over2} + 2\times n,
\eqno(2.18)
$$
respectively. The $R^{ab}$ and $R_{ab}$ generators again form the rank-2
antisymmetric representation, but now we also have rank-1 contributions
from $R^a$ and $R_a$. These objects transform as vectors and
co-vectors under $SL(n)$. The $R$-generators with upper and lower
indices belong to the positive and negative roots of $B_n$
respectively.  Given our choice of local subalgebra  the positive root
generators lead to Goldstone bosons in the non-linear realisation.
\par
We show next, that we can in fact find the Borel subalgebra of $B_8$
which has the  commutation relations
$$
   [K^a{}_b,\, K^c{}_d] = \delta _b^c\, K^a{}_d - \delta^a_d\, K^c{}_d
$$
$$
   [K^a{}_b,\, R^c] = \delta_b^c\, R^a,\quad [K^a{}_b,\, R^{cd}] =
   \delta^c_b \, R^{ad} + \delta^d_b R^{ca},
$$
$$
   [R^a,\, R^{bc}] =0,\quad [R^a,\, R^b] = -R^{ab}\quad ({\it here:}\,
   {\rm all\,\,indices \,\, 1,\ldots,8})
\eqno(2.19)
$$
from the commutation relations of $G_{Ia}$ after a suitable rescaling.
We start by allowing seven values for the indices $i,j = 4,\ldots,
10$. The new simple positive root is called $E_4 = K^4{}_5$, the new
Cartan element has to be $H_4 = K^4{}_4 - K^5{}_5$ in order to find the
correct $SL(7,\, {\bf R})$ subgroup (the commutators with the positive
root of $A_1$ and $E_{10}$ ensure that there are no trace or dilaton
parts $\hat D$ or $R$). We note that $[H_4,\, E_4] =2\, E_4$ as it
must be, and $[H_4,\, E_5] =-E_5$. We also find $[H_4,\, E_{-1}] =
-E_{-1}$, which tells us that the new node couples the extra node
$A_1$ to the existing $B_6$ diagram to give $B_8$. We define the
generators
$$
   \hat K^{-1}{}_i = {1\over 6!} \epsilon_{ii_1\ldots i_6}
   R^{i_1\ldots i_6}
\eqno(2.20)
$$
and by dualising the seven-form generator we extend the vector
generator:
$$
   \hat R^{-1} = - {7\over \sqrt{2}\, 7!} \epsilon_{i_1\ldots i_7}
   R^{i_1\ldots i_7}.
\eqno(2.21)
$$
We seem to have run out of Goldstone fields in the theory to also
enhance the $R^{ab}$-generator. As realised previously, we must
introduce duals of gravity to find the missing Goldstone fields (see
first chapter). This happens by altering the existing commutation
relations (2.4) to now become
$$
   [R^{c_1\ldots c_6},\, R^{bc}] = -c_{2,6} R^{c_1\ldots c_6bc} +
   {2\over 7} R^{c_1\ldots c_6[b,c]}.
\eqno(2.22)
$$
The Jacobi identities then imply the relation
$$
   [R^{c_1\ldots c_7},\, R^b] = -c_{1,7} R^{c_1\ldots c_7b} +
   {2\over 7} R^{c_1\ldots c_7,b}.
\eqno(2.23)
$$
This new generator, corresponding to dual of gravity, can in seven
dimensions be dualised and used to define
$$
   \hat R^{-1\,j} = {1\over 7!} \epsilon_{i_1\ldots i_7} R^{i_1\ldots
   i_7,j}.
\eqno(2.24)
$$
We also have to introduce a new generator for the Cartan subalgebra
$$
   \hat K^{-1}{}_{-1} = a\, \hat D + b \, R,\quad \hat D =
   \sum_{l=4}^{10}K^l{}_l.
\eqno(2.25)
$$
The constants can be fixed from the commutation relations with other
generators. In particular we require
$$
   [\hat K^{-1}{}_{-1},\, R^{ij}] =0,\quad [\hat K^{-1}{}_{-1},\, \hat
   K^{-1}{}_j] = \hat K^{-1}{}_j,\quad[\hat K^{-1}{}_{-1},\, \hat
   R^{-1\,i}] = \hat R^{-1\,i}
\eqno(2.26)
$$
and thus find that  $a= {1\over 8}$ and $b= -{1\over 2}$. We have no
degrees of freedom left, but it is a non-trivial check that for these
values we yield the right commutation relations for the vector
$$
   [\hat K^{-1}{}_{-1},\, \hat R^i] = 0,\quad [\hat K^{-1}{}_{-1},\,
   \hat R^{-1}] = \hat R^{-1},
\eqno(2.27)
$$
as indeed we do. Using the slightly modified generators
$$
   \hat K^i{}_j = K^i{}_j - {1\over 8}\, \hat D + {1\over 2}\, R =
   K^a{}_b - \hat K^{-1}{}_{-1},
\eqno(2.28)
$$
one can check that these generators indeed fulfill the correct
commutation relations of the Borel subalgebra of $B_8$ when written in
terms of representations of $SL(8,\, {\bf R})$ as given in equation
(2.19) if the eight indices are taken to be $i,j = -1,\,4,\ldots,
10$.
\par
Since $G_{Ia}$ is a symmetry of the supergravity theory with one
abelian vector it follows that also the Borel subalgebra of $SO(8,9)$
is a symmetry once we have found a formulation involving the duals of
gravity. A theory invariant under the full $B_8$ will have to include
the symmetries under the negative roots $R_{ab}$ and $R_a$. These
symmetries are understood to be unbroken local symmetries and therefore
do not contribute towards the Goldstone spectrum. They enhance the local
Lorentz symmetry ($SO(8)$) and should correspond to hidden symmetries of
the  equations of motion.
\par
We know that the ten dimensional theory must have $SL(10,\, {\bf R})$ as
a symmetry group too.  We summarise that the Kac-Moody group we
want  to identify needs to have $B_8$ and $A_9$ as subalgebras when
accordingly restricted. The algebra of $B_8$ occurred when the indices
assumed the seven values $i,j = 4,\ldots, 10$. It naturally included
(see 2.19) the $K^i{}_j$ generators of the $A_6$ subalgebra. We now
have to enhance this $A_6$ subalgebra to the full $A_9$ subalgebra and
clarify how the additional three nodes couple to the rest of the
Dynkin diagram. Looking at the positive roots (2.14) of the $G_{Ia}$
algebra, and in particular at the simple positive roots (2.16), we
find that the generators corresponding to the  vector potential and
6-form potential transform as rank one and six anti-symmetric tensors
under the $A_9$ subalgebra (2.3). The commutations of equations (2.3)
imply that the 6-form generator $R^{a_1\ldots a_6}$ and the vector
generator $R^a$ transform  as their indices indicate under
$A_9$. Using the discussion  from the first chapter
concerning  the addition of extra nodes to the Dynkin diagram
of $A_n$, we conclude that  the corresponding simple generators they
contain  can only   attach respectively to the first and sixth nodes from
the right of the $A_{9}$ line. In particular, there are no additional
connections between the nodes $1,\,2,\,3$ and the nodes
$-1,\,10$. This resulting group is very-extended $B_8$, whose Dynkin
diagram is given in {\it Figure C}. We also note that very-extended
$B_8$ contains affine $B_8$ if the indices do not assume values
$1,\,2$.

\vbox{
\vskip.2cm
\hbox{\hskip3cm\epsfbox{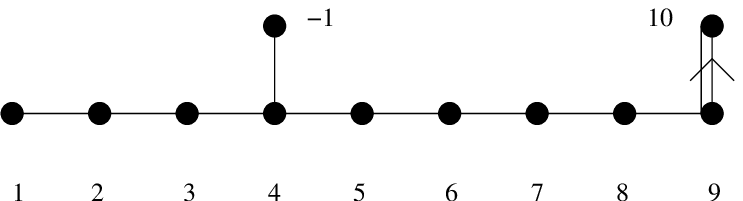}}
\vskip.2cm}
\hskip2cm {\it Figure C: }{The Dynkin diagram of very-extended $B_8$}
\vskip.5cm
\noindent
The Chevalley generators of the Kac-Moody algebra have as their
simple positive roots
$$
   E_a = K^a{}_{a+1}\quad {\rm for}\quad a =1,\ldots, 9,\quad E_{10}=
   R^{10},\quad E_{-1} = R^{56789\,10}
\eqno(2.29)
$$
as well as the Cartan sub-algebra elements
$$
   H_a = K^a{}_a - K^{a+1}{}_{a+1}\quad {\rm for}\quad a =1,\ldots,
   9,\quad H_{10}= 2 K^{10}{}_{10} - {1\over 4}\, D + {1\over 2}
   \,R,
$$
$$
   H_{-1} = - K^1{}_1 - K^2{}_2 - K^3{}_3 - K^4{}_4 + 2
   K^{-1}{}_{-1},
\eqno(2.30)
$$
where the unhatted $D = \sum_{l=1}^{10} K^l{}_l$ runs over all ten
values (also in the definition of $K^{-1}{}_{-1} = {1\over 8} D -
{1\over 2}R$). Except for $H_{10}$, and $E_{10}$ these generators
coincide with those of pure supergravity described in the previous
chapter.


%


%

\medskip
{\bf 3. $N=1$ supergravity with more than one vector multiplet}
\medskip
The classification of semi-simple Lie algebras is usually carried out
when the algebras are taken to be over the complex  numbers. Consequently,
the end result of the classification, i.e. the Dynkin diagram, does not
specify a preferred real form of the algebra. Any finite dimensional
complex Lie algebra possesses a unique  real form  in  which all the
generators are compact. It is given by taking the generators
$U_\alpha=i(E_\alpha+E_{-\alpha})$, $V_\alpha=(E_\alpha-E_{-\alpha})$
and $iH_{a}$, where $\alpha$ is any positive root. The compact nature of
the generators follows in an obvious way form the fact that
$(E_\alpha, E_{-\alpha})>0$. By considering all involutive automorphisms
of the unique compact real form one can construct all other real forms
of the complex Lie algebra under consideration. In particular, the real
forms are in one to one correspondence with all those involutive
automorphisms of the compact real algebra which can not be written as
real transformations.
\par
Given an automorphism ${\cal I}$ which is involutive (${\cal{I}}^2=1$) we
can divide the generators of the compact real form into those which
possess $+1$ and $-1$ eigenvalues of ${\cal I}$. We denote these
eigenspaces by
$$
    \cal{G} = \cal{K} \oplus \cal{P}
\eqno(3.1)
$$
respectively. Since ${\cal I}$  is an automorphism, the algebra when
written in terms of this split takes the generic form
$$
    [{\cal{K}},\,{\cal{K}}] \subset {\cal{K}},\quad [{\cal{K}},
    {\cal{P}}] = {\cal{P}},
\qquad  [{\cal{P}},\, {\cal{P}}] \subset
    {\cal{K}}.
\eqno(3.2)
$$
For the generators ${\cal P}$ we define new generators
$\hat{\cal P}=i{\cal P}$, hence the algebra takes the generic form
$$
    [{\cal{K}},\,{\cal{K}}] \subset {\cal{K}},\quad [{\cal{K}},\,
    \hat {\cal{P}}] = \hat {\cal{P}},\quad [\hat {\cal{P}},\, \hat
    {\cal{P}}] \subset (-1) {\cal{K}}.
\eqno(3.3)
$$
Thus we find a new real form of the algebra in  which the generators
${\cal K}$ are compact while the generators $\hat {\cal P}$ are
non-compact. This follows from the fact that all the generators in
the original algebra are compact and so have negative definite scalar
product. Hence, the number of non-compact generators changes from one
real form to another.
\par
An important real form can be constructed by considering the Cartan
involution which is a linear operator that takes
$E_{\alpha}\leftrightarrow -E_{-\alpha}$ and $H_a \to -H_a$. Clearly
the generators of the compact real form transform as
$V_\alpha\  \to V_\alpha$, $U_\alpha\  \to -U_\alpha$ and $H_a\ \to
-H_a$. Using this involution we find a real form with generators
$$
   \hat V_\alpha=V_\alpha=E_\alpha-E_{-\alpha},\
   \hat U_\alpha=-iU_\alpha=E_\alpha+E_{-\alpha} ,\ \hat H_a=-i H_a.
\eqno(3.4)
$$
The $\hat V_\alpha$ remain compact generators while $\hat U_\alpha$ and
$\hat H_a$ become non-compact. The maximal compact sub-algebra is just
that invariant under the Cartan involution. The real form of the
algebra found in this way has the maximal number of non-compact
generators of all the real forms one can construct. It is therefore
called the maximal non-compact real form. For  example, the complex Lie
algebra $D_{n}$ has $SO(2n)$ as its unique compact real form and $SO(n,n)$
as its maximal non-compact real form. For a more detailed discussion of
real forms see [17].
\par
Given a real form of a complex Lie algebra, any element $g$  of the
associated group can be expressed as $g=g_c g_{na} g_r$ where
$g_c$ is in the maximal compact sub-group, $g_{na}$ is the maximal
commuting non-compact subalgebra and $g_r$ is the group found by
exponentiating the generators which are in the positive root space of
$g_{na}$. For the case of a maximally non-compact form of the algebra,
all the Cartan generators are non-compact and so $g_{na}$ is just
the Cartan sub-algebra while $g_r$ is generated by all positive root
generators of the original algebra. As such this is the decomposition
for which $ g_{na} g_r$ is just the Borel sub-group. The important
point to note is that it is only for the maximally non-compact real
form that all Cartan subalgebra elements of the original algebra appear
in $g_{na}$ and all simple roots can be generated from multiple
commutators of the generators appearing in $g_r$.
\par
The construction of non-linear realisations based on a given algebra
is carried out with respect to a particular real form of a given
algebra. An important ingredient in the construction is the choice of
local subalgebra as this effects the field content and the way the
symmetries are realised. The local subalgebra is usually chosen to be
the maximally compact subalgebra of a the real form being used. Clearly,
this subalgebra changes from one real form to another. For example, for
$SO(n,n)$ the maximally compact sub-algebra is $SO(n)\otimes SO(n)$ while
for $SO(p,q)$ it is $SO(p)\otimes SO(q)$. Clearly , even if $p+q=2n,
p\not=q$ the dimensions of the two cosets are different and so is the
physics resulting from the two  non-linear realisations based on the two
algebras. In particular, only for $SO(n,n)$ do all the Cartan subalgebra
generators appear in the coset. So far, all algebras considered in the
context of the  eleven dimensional supergravity, IIA and IIB supergravity
and the $N=1$ supergravity theory coupled to no or one vector multiplet
were the maximal non-compact form of real algebras. However, we will see
that with more than one vector multiplet coupled to $N=1$ supergravity
one must consider symmetry algebras that are not the maximally
non-compact real form.
\par
The field contents of $N=1$ supergravity with an arbitrary number of
abelian vectors $n$ added is given by
$$
   h_a{}^b,\quad A,\quad A^{(k)a},\quad A^{a_1a_2},\quad A^{a_1\ldots
   a_6}, \quad A^{(k)a_1\ldots a_7},\quad A^{a_1\ldots a_8},
\eqno(3.5)
$$
where the indices $a_i,\, b$ take ten values according to the ten
dimensional tangent space, while the index $(k)$ numerates the abelian
vectors $1,\ldots, n$. Since we wish to express the theory as a
non-linear realisation we introduce the corresponding generators
$$
   K^a{}_b,\quad R,\quad R^{(k)}_a,\quad R_{a_1a_2},\quad R_{a_1\ldots
   a_6},\quad R^{(k)}_{a_1\ldots a_7},\quad R_{a_1\ldots a_8}.
\eqno(3.6)
$$
We can define a group element in total analogy to the preceding cases
$$
   g = \exp (x^\mu\, P_\mu)\, \exp(h_a{}^b\, K^a{}_b)\,g_A \equiv g_x\,
   g_h\, g_A,
\eqno(3.7)
$$
where now
$$
   g_A = \exp({1\over 8!} A_{a_1\ldots a_8} R^{a_1\ldots a_8})\,
   \exp({1\over 7!}\sum_k\,A^{(k)a_1\ldots a_7} R^{(k)}_{a_1\ldots
   a_7}) \exp({1\over 6!} A_{a_1\ldots a_6} R^{a_1\ldots a_6})
$$
$$
   \times \exp({1\over2} A_{a_1a_2} R^{a_1a_2})\, \exp(\sum_k \,
   A^{(k)}_a R^{(k)a}  ) \exp(AR).
\eqno(3.8)
$$
We take the commutation relations
$$
  [K^a{}_b,\, K^c{}_d] = \delta^c_b K^a{}_d - \delta^a_d K^c{}_b,\quad
  [K^a{}_b,\, P_c] = -\delta^a_c P_b,
$$
$$
   [K^a{}_b,\, R^{a_1\ldots a_p}] = \delta^{a_1}_b R^{aa_2} +
   \ldots,\quad [R,\, R^{a_1\ldots a_p}] = c_p R^{a_1\ldots a_p},
$$
$$
   [R^{a_1a_2},\, R^{b_1\ldots b_6}] = c_{2,6} R^{a_1a_2b_1\ldots b_6},
\eqno(3.9)
$$
where the index $(k)$ which turns up in the second line for some
generators is suppressed as it is unaffected by the action of $K^a{}_b$,
as well as the commutation relations
$$
   [R^{(k)a},\,R^{(l)b}] = c_{1,1}\delta^{k,l}\, R^{ab},\quad
   [R^{(k)a},\,R^{a_1\ldots a_6}] = c_{1,6}R^{(k)a_1\ldots a_7}
$$
and
$$
   [R^{(k)a},\,R^ {(l)a_1\ldots a_7}] = - c_{1,7} \delta^{k,l}\,
   R^{aa_1\ldots a_7}.
\eqno(3.10)
$$
The constants are as in (2.4), and $\delta^{(kl)}$ is equal to one
if $k=l$ and zero otherwise. The algebra defined by the commutation
relations in (3.9), (3.10) we call $G_{IA(n)}$, where the subscript
reminds us of the fact that we are still dealing with $N=1$ abelian
theories, however, this time with an arbitrary number of vectors $n$.
\par
Proceeding as usually by calculating the covariant forms of the
non-linear realisation and taking only those that are also covariant
under the conformal group we derive the following field strengths which
are very similar to those of ten dimensional supergravity with one
vector multiplet, just enhanced by an additional index:
$$
   \tilde F_a = \tilde D_a A,
\eqno(3.11)
$$
$$
   \tilde F^{(k)}_{a_1a_2} = 2 e^{-1/4 A} \tilde D_{[a_1}
   A^{(k)}_{a_2]},
\eqno(3.12)
$$
$$
   \tilde F_{a_1a_2a_3} = 2 e^{-1/2 A} (\tilde D_{[a_1} A_{a_2a_3]} - 2
   A^{(k)}_{[a_1}\tilde D_{a_2} A^{(k)}_{a_3]})
\eqno(3.13)
$$
for the original fields, and for the duals of those fields we find
$$
   \tilde F_{a_1\ldots a_7} = 7 e^{1/2 A} \tilde D_{[a_1} A_{a_2\ldots
   a_7]},
\eqno(3.14)
$$
$$
   \tilde F^{(k)}_{a_1\ldots a_8} = 8 e^{1/4 A} (\tilde D_{[a_1}
   A^{(k)}_{a_2\ldots a_8]} - 7A^{(k)}_{[a_1}\tilde D_{a_2}
   A_{a_3\ldots a_8]})
\eqno(3.15)
$$
and
$$
   \tilde F_{a_1\ldots a_9} = 9 (\tilde D_{[a_1} A_{a_2\ldots
   a_9]} - 2\cdot 7A_{[a_1a_2}\tilde D_{a_3} A_{a_4\ldots a_9]} + 4
   A^{(k)}_{[a_1}\tilde D_{a_2} A^{(k)}_{a_3\ldots a_9]} ),
\eqno(3.16)
$$
where a sum over the $(k)$ indices is understood. Although the addition
of vectors changes the theory considerably (anomalies), in terms of
field strengths and field equations the change is marginal; the only
covariant field equations are similarly to (2.12) given by
$$
   \tilde F^{a_1\ldots a_p} = {1\over (10-p)!} \epsilon^{a_1\ldots
   a_{10}} \tilde F_{a_{p+1}\ldots a_{10}}, \quad {\rm for}\,\, p
   =1,3.
$$
and
$$
   \tilde F^{(k)a_1a_2} = {1\over 8!} \epsilon^{a_1\ldots
   a_{10}} \tilde F^{(k)}_{a_{3}\ldots a_{10}}.
\eqno(3.17)
$$
These field equations are the correct field equations for the bosonic
sector of $N=1$ supergravity coupled to $n$ abelian vector multiplets.
\par
We now consider how the above algebra may be embedded  into a known
Kac-Moody algebra. Again we omit the momentum generators and begin by 
splitting the other generators into three different classes:
$$
   G^+_{IA(n)} = \{ K^a{}_b,\, a<b,\, a,b=1,\ldots, 10,\quad R^{(k)a},\,
   R^{a_1a_2},\,R^{a_1\ldots a_6},\, R^{(k)a_1\ldots a_7},\,
   R^{a_1\ldots a_8}\},
\eqno(3.18)
$$
and
$$
   G^0_{IA(n)} = \{ H_a = K^a{}_a - K^{a+1}{}_{a+1},\, a= 1,\ldots,9,\, D=
   \sum_a K^a{}_a,\quad R\}
\eqno(3.19)
$$
plus $K^a{}_b$ where $a>b$.
The elements of $G^+_{IA(n)}$ can be generated by multiple commutators of
the generators
$$
   K^a{}_{a+1} \quad {\rm for} \quad a = 1,\ldots, 9,\quad
   R^{(k)10},\ k=1,\ldots ,n,\ \quad {\rm and}\,\, R^{56789\,10}.
\eqno(3.20)
$$
In the cases previously considered by the authors we have identified
the generators analogous to those in equation (3.20) as the simple
root generators and those in equation (3.19) as elements of the Cartan
sub-algebra of the proposed Kac-Moody algebra. However, unlike in the
cases previously considered, the number of elements in equations (3.20)
and (3.19) does not match except if $n=1$, which was considered in the
last section, and so this identification can not be quite right if $n>1$.
\par
The resolution of this dilemma is that for $n>1$ we are not dealing with
the maximally non-compact real form of an  algebra. Thus the local
sub-algebra -the maximally compact sub-algebra- is different. In the
non-split case, it too contains elements of the Cartan sub-algebra.
We illustrate the point by considering the algebra $SO(8,\, 8+n)$ whose
maximal compact sub-algebra is $SO(8)\otimes SO(8+n)$. It is useful to
decompose $SO(8,\, 8+n)$ into representations of $SL(8)\otimes  SO(n)$.
The commutators of the resulting generators are as follows:
$$
   [K^a{}_b,\, K^c{}_d] = \delta _b^c\, K^{}_d - \delta^a_d\, K^c{}_d,
   \quad [K^a{}_b,\, R^{(l)c}] = \delta_b^c\, R^{(l)a},\quad [K^a{}_b,\,
   R^{cd}] = \delta^c_b \, R^{ad} + \delta^d_b R^{ca},
$$
$$
   [R^{(l)a},\, R^{bc}] =0,\quad [R^{(k)a},\, R_{(l)}^b] =
   -\delta^k_l\, R^{ab}, \quad [R^{(k)}_a,\, R_{(l)b}] =
    \delta^k_l\, R_{ab},
$$
$$
   [K^a{}_b,\, R^{(l)}{}_c] = - \delta^a_c\, R^{(l)}{}_b,\quad [K^a{}_b,\,
   R_{cd}] = - \delta^a_c \, R_{bd} - \delta^a_d R_{cb},
$$
$$
   [S^{kl},\, R^{(m)a}] = \delta^{(km)}\, R^{(l)a} - \delta^{(lm)}\,
   R^{(k)a} ,\quad [S^{kl},\, R^{ab}] = 0 = [K^a{}_b,\, S^{kl}],
$$
$$
   [R^{ab},\, R^{(k)}_c] = \delta^{[a}_c\, R^{(k)b]},\quad [R_{ab},\,
   R^{(k)c}]= \delta_{[a}^c\,R^{(k)}_{b]},
$$
$$
   [R^{(k)a},\, R^{(l)}_b] = \delta^{(kl)} K^a{}_b + \delta^a_b S^{kl},
   \quad [R^{ab},\, R_{cd}] = \delta^{[a}_{[c} \, K^{b]}{}_{d]}.
\eqno(3.21)
$$
where the indices take the values $a,b,c=  1,\ldots,8$ and
$k,l,m = 1,\ldots, n$), and $\delta^{(kl)} =1$ if $k=l$ and zero
otherwise. The position of the $(k)$ index distinguishes between vector
and covector with respect to the $SO(n)$ subalgebra. They are simply
transposed to each other and thus their entries coincide $R^{(l)a}=
R^a_{(l)}$ and $\delta^{(kl)}=\delta^k_l$.
\par
The generators of the maximal compact sub-algebra among (3.21) are given
by the anti-symmetric part in $2K_{[ab]}= J_{ab}$ which generate $SO(8)$,
and the linear combinations $R_{ab} + R^{ab}$, $R^{(k)}_a- R^{(k)a}$ and
$S^{kl}$ which can easily be shown to generate $SO(8+n)$.
\par
Taking the maximal compact subalgebra, i.e. $SO(8)\otimes SO(8+n)$, as
the local sub-algebra in the non-linear realisation, the coset is of the
form $g_{na}g_r$ and contains the generators $K^{a}{}_b, a\le b$,
$R^{ab}$, $R^{(k)a}$. We divide the coset generators into a commuting
set given by $K^{a}{}_{a}-K^{a+1}{}_{a+1},\  a= 1,\ldots ,7 $ as well
as a trace term $D=\sum_1^8 K^a{}_a$, and the remainder, which can be
expressed as multiple commutators of $K^{a}{}_{a+1},\ a= 1,\ldots ,7 $
and $R^{(k) 10}$. We note that the commuting set contains only 8
generators while the remainder is generated by $7+n$ elements,
where $n$ is the number of added vectors and thus $n>1$ in this chapter.
Hence we have a mismatch identical to that found above. Indeed,
as explained there some of the Cartan sub-algebra generators of the
original algebra appear in the subgroup and indeed all the generators
$S^{kl}$ are in the sub-algebra.
\par
Having understood the consequences of taking a real form that is not the
maximally non-compact real form, we now continue with identifying  the
Kac-Moody algebra. Let us start by considering  the case of only two
vector multiplets, i.e. $n=2$. This does not effect the counting in
(3.19), so we still find eleven commuting elements in the coset.
However, we  find now (3.20) that 12 elements are needed to generate all
generators of the non-linear realisation in (3.6) via repeated
commutation relations. To identify the underlying algebra, we again
restrict the space-time indices to take only seven values $4,\ldots,10$,
and define
$$
   \hat K^{-1}{}_i = {1\over 6!} \epsilon_{ii_1\ldots i_6} R^{i_1\ldots
   i_6} ,
$$
$$
   \hat R^{(k)-1} = {1\over \sqrt{2}\, 7!}\epsilon_{i_1\ldots i_7}
   R^{(k)i_1\ldots i_7} ,\quad k=1,2
$$
$$
   \hat R^{-1\,i} = {1\over 7!}\epsilon_{i_1\ldots i_7}
   R^{i_1\ldots i_7,i}.
\eqno(3.23)
$$
In the first line, the six-form generator was dualised and used to define
a $\hat K^{-1}{}_i$ (index enhancement by one), in the second line the
duals of the gauge fields were put into the original gauge field generator
(we again obtain an index enhancement, this time for the vectors). In the
last equation, the  generator corresponds to a field dual of gravity.
This generator can be obtained from altering the following commutators
in $G_{IA(n)}$
$$
   [R^{a_1\ldots a_6},\, R^{b_1b_2}] = -c_{2,6} R^{a_1\ldots
   a_6b_1b_2} + {2\over 7} R^{a_1\ldots a_6[b_1,b_2]},
$$
$$
   [R^{(l)a_1\ldots a_7},\, R^{(k)b}] = -c_{2,6}\delta^{(kl)}\,
   R^{a_1\ldots a_6b_1b_2} + {2\over 7}\sum \delta^{(kl)}
   R^{a_1\ldots a_7,b}.
\eqno(3.24)
$$
These commutation relations  fulfill the Jacobi identities, and
they coincide with the commutation relations of the generators of the
subalgebra of (3.21) for the case $n=2$, i.e. $SO(8,10)$ which are not
in the maximally compact sub-algebra. We do not expect to find the
generators of the compact sub-algebra as these do not lead to fields
in the non-linear realisation since they belong to the local symmetries.
Deploying similar arguments to those given in the previous chapters we
can argue that the underlying Kac-Moody algebra which allows for the
full index range $a,b =1,\ldots,10$ must be very-extended $D_9$ whose
real form is very extended SO(8,10).
\par
We now identify the Cartan sub-algebra and positive simple roots
generators of  very-extended $SO(8,8+n)$. We have just noted that when
the group involved is not split some of the generators of the Cartan
sub-algebra are compact and some are non-compact. Clearly, the compact
generators of the Cartan sub-algebra belong to the compact subalgebra and
so do not appear explicitly in the non-linear realisation. Using
elements of the Cartan sub-algebra of the compact subalgebra $SO(n)$
generated by $S^{kl}$, one can easily check that for each value of $n$
the number of Cartan generators from this compact subalgebra accounts
precisely for the mismatch between maximally commuting non-compact
generators (3.19) and the generators that span the positive roots
of the coset elements (3.20).
\par\noindent
All Cartan generators of the full very-extended algebra in the case of
two added vectors, i.e. an $SO(2)$ subgroup with generator $iS^{12}$,
are uniquely given by
$$
    H_a = K^a{}_a - K^{a+1}{}_{a+1} \quad {\rm for} \quad a = 1,
   \ldots,9, \quad H_{10} = K^{10}{}_{10} + iS^{12} + {1\over 8}D
   - {1\over 2} R,
$$
$$
   H_{11} = K^{10}{}_{10} - iS^{12} - {1\over 8} D
   + {1\over 2} R,\quad H_{-1} = -K^1{}_1- K^2{}_2 - K^3{}_3 - K^4{}_4 +
   {1\over 4} D - R.
\eqno(3.25)
$$
We note that the linear combination defined in chapter one (1.35) and
two (2.30) by $K^{-1}{}_{-1}= {1\over 8}D -{1\over 2}R$ can again be
made use of in order to simplify the expressions for $H_{10}$, $H_{11}$,
and $H_{-1}$.
\par
We also want to define the simple positive roots which together with
(3.25) can be used to derive the complete Dynkin diagram of the
Kac-Moody algebra that is called very-extended $D_9$. They have to
meet two requirements: they have to be eigenvectors of the Cartan
sub-algebra elements (with respect to the adjoint representation),
and they have to be linear combinations of the simple elements found
in (3.20). As such they can only be
$$
   E_a=K^a{}_{a+1}\quad{\rm for}\quad a=1,\ldots,9,\qquad
      E_{-1}=R^{56789\,10}
$$
$$
   E_{10}=R^{(1)10}+ iR^{(2)10},\quad E_{11}=R^{(1)10}-iR^{(2)10}.
\eqno(3.26)
$$
The occurrence of the imaginary unit is related to the fact that we
are in this case dealing with non-split forms of real groups. 
\par
We will show next how the above two vector generators correspond to
new nodes in the resulting Dynkin diagram. We restrict our attention to 
the finite Lie sub-algebra the extension to the  very extended
algebra being obvious. 

\input epsf.tex
\vbox{
\vskip.2cm
\hbox{\hskip3cm\epsfbox{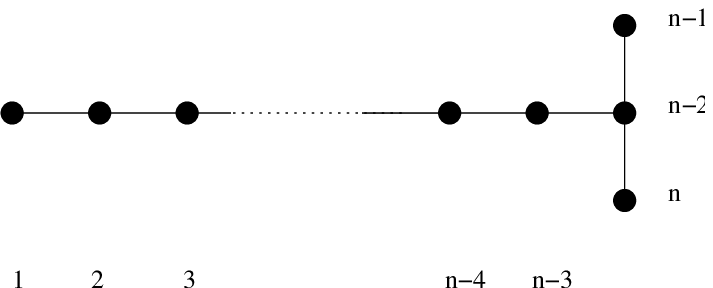}}
\vskip.2cm}

\hskip2cm {\it Figure D: }{The Dynkin diagram of the $D$-series}
\vskip.5cm
\noindent
By this way of drawing the Dynkin diagram it is understood, that $D_n$
is decomposed into representations of $A_{n-2}$ (the horizontal line)
rather than $A_{n-1}$. Decomposition into representations of $A_{n-2}$
gives 2  vector generators, apart from the known 2-form
generator. The decomposition of $SO(n-1,n+1)$ into representations of
$SL(n-1)\times SO(2)$ splits the $n(2n-1)$ degrees of freedom of the
former into representations $K^a{}_b$, $D$, $R^{ab}$ and $R_{ab}$,
$R^{a(i)}$ and $R_{a(i)}$ (where $a,b =1,\ldots, n-1$), plus the
$SO(2)$ generator $iS^1{}_2$. The two added vectors effectively span
the new direction (called 11 in {\it Figure E}). The resulting
symmetry algebra is thus $D_9$, in the real form $SO(8,10)$. 
\par
The generalisation to an arbitrary number $n$ of abelian vectors should
follow the same pattern since the above discussion was
not special to the case of only two added vectors. In particular, one can
easily allow for arbitrary $k$ in the second line of equation (3.23)
where we restricted the index range to seven space-time dimensions. Every
pair ($l$) of vectors ($E_{8+2l}= R^{(2l-1)\,10} + iR^{(2l)\,10}$ and
$E_{9+2l}= R^{(2l-1)\,10} - iR^{(2l)\,10}$ for $l=1,2,3,\ldots$) will
just increase the rank of the algebra. If the number of vectors is even
($n=2k$), then the resulting algebra will be $D_{8+k}$ in the real form
$SO(8,8+n)$, if the number of vectors is odd ($n=2k+1$), then the
resulting  algebra will be $B_{8+k}$ in the real form $SO(8,8+n)$,
where the last vector stays without partner. The algebra that mixes the
different flavours of the abelian vectors is $SO(n)$. The addition of
vectors follows the embedding
$$
   D_8\subset B_8\subset D_9\subset B_9 \subset \ldots,
\eqno(3.27)
$$
where the leftmost group corresponds to pure supergravity from chapter
one, the second entry to one vector as described in chapter 2. This
embedding also holds for the very-extended versions of those groups.
In terms of Dynkin diagrams the extension works like in {\it Figure E}.

\vbox{
\vskip1cm
\hbox{\hskip3cm\epsfbox{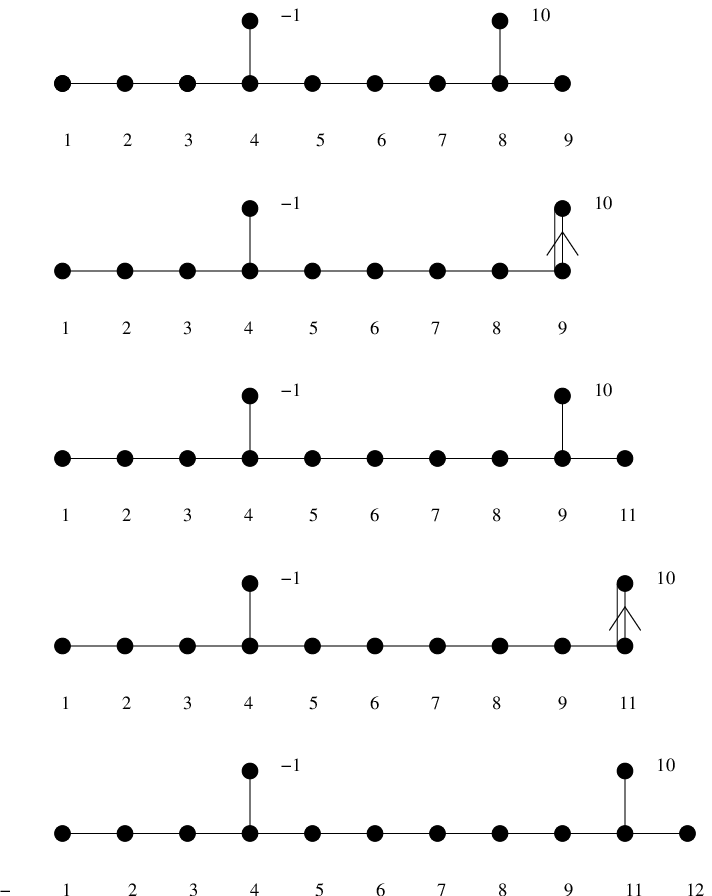}}
\vskip.2cm}
\hskip1cm {\it Figure E: }{Growing Dynkin diagrams for added vectors}
\vskip.5cm
\noindent
We have called the extra dimension that showed up after adding the first
two vectors ``11'', and the extra dimension for 4 added vectors, we have
called ``12'' in the above {\it Figure E}.
\par
We notice that adding 32 vectors to the theory will give the Kac-Moody
algebra of very-extended $D_{24}$. This algebra was also found to be
the symmetry algebra of the bosonic string [11]. Interestingly, the new
dimensions in our way of labeling would be called $11,\ldots, 26$.
Hence, we find the same Dynkin diagram for a 10-dimensional theory if
we add in 32 vectors! However, the Kac-Moody algebra $D_{24}^{+++}$ for 
the effective bosonic string uses a different real form which for the 
finite dimensional sub-algebra is $O(24,24)$. In the heterotic case, we 
required the real form $O(8,40)$. Although the Dynkin diagrams of 
different real forms coincide because of the same complex algebra they 
are derived from, we require different real symmetry groups to obtain the 
correct field content and Lorentz symmetry for each theory. Furthermore, 
as we have considered only abelian vectors, the $SO(32)$-symmetry that we 
find  is not a gauge symmetry, and there are  no gauge connection terms in 
the field strengths $F^k_{a_1a_2}$, as can be seen from equation (3.12). 
Nonetheless, this result goes in the direction of the conjecture that all 
string theories can be derived from the closed bosonic string [18]. 
\par 
We close this section with some comments on how one might gauge the
$SO(32)$ symmetry. Since gravity reduced on a sphere leads to
non-abelian gauge symmetries and gravity can be described in an arbitrary
background by a non-linear realisation whose underlying Kac-Moody algebra
is thought to be $A_{D-3}^{+++}$ [19] it must be possible to include
non-abelian vector fields in the framework developed in this paper. 
Indeed,  there is a close analogy between the way the dual fields for
gravity and the dual gauge fields occur in the Kac-Moody algebra. The
corresponding generators are $R^{a_1\ldots a_7,b}$ and 
$R^{(l)a_1\ldots a_7}$ respectively and these generators  both first 
occur when considering the restriction of the the index range to take 
seven values  ($4,\ldots,10$) since their index structure is fairly 
similar. 
\par
We note that if we allow the derivatives of the gauge fields to
become covariant derivatives then the equation 
$$
   D_a F^{(k)ab} = {1\over 2!7!}\epsilon^{ba_1\ldots a_9}\,
    H_{a_1\ldots a_7}\, F^{(k)}_{a_8a_9}.
\eqno(3.28)
$$
is indeed the correct field equation
for the gauge vectors of heterotic supergravity in 10 dimensions. It is
just the non-abelian analogue of the abelian field equation given in
(3.17).

%
%

\medskip
{\bf 4. Discussion}
\medskip
In this paper we have found that the bosonic sector of the $N=1$
supergravity theories in ten dimensions coupled to any number $n$ of
abelian multiplets can be described by a non-linear realisation and we
have argued that this theory can be extended in such a way that it
possess a Kac-Moody algebra that is $D_{8+{n\over 2}}$ if $n$ is even 
and $B_{8+{n-1\over 2}}$ if $n$ is odd. However, the construction uses 
particular real forms which for the finite dimensional sub-algebras are 
$O(8,8+n)$. 
\par
These underlying Kac-Moody algebras are consistent with a number of
related results in the literature. It has been known [20] for many years
that the supergravity system mentioned above when dimensionally reduced 
to $d,\,  d\ge 4$ space-time dimensions possesses scalars which are in  
a non-linear realisation of the algebra $O(d,d+n)$ with local subgroup 
$O(d)\times O(d+n)$. The symmetry algebra  in lower dimensions can be 
read off from the Kac-Moody Dynkin diagram of the theory in ten 
dimensions by deleting nodes from the left. Carrying out this procedure, 
we indeed find the  symmetry algebra of reference [20]. 
\par
It has been found that the dynamics of certain theories of gravity
coupled to a dilaton and $n$-forms near a space-like singularity becomes   
a one dimensional  motion with scattering taking place in the Weyl chambers 
of certain overextended Lie algebras [21]. This motion is sometimes   
referred to as cosmological billiards.  For $N=1$ pure supergravity
theory the corresponding algebra that appears is over-extended $D_8$ 
(denoted $D_8^{++}$) and for this theory extended with a vector field is
over-extended $B_8$ (denoted $B_8^{++}$) [21]. 
\par
It has also been shown [22] that the $T$-duality transformations for the 
IIA string theory  in ten dimensions reduced on an $k$-torus for 
$k=1,\dots,10$  have a natural action on the moduli space of the $k$-torus 
that is   the Weyl group of $E_k$. The corresponding result for the 
supergravity with 32 vector fields reduced to one dimensions is the Weyl 
group of over-extended O(8,24) [23].  
\par
The existence of these huge symmetry groups in gravity theories must 
be related to the symmetries of solutions to these theories, and might 
thus be utilised to find solutions of the supergravity theories [24]. 
More importantly, it has been suggested [10-16,19,25,26] that 
the non-linear realisation constructed from any very-extended algebra 
$G^{+++}$, denoted ${\cal V}_{\cal G}$  in [25], can provide a 
consistent theory  which extends that of gravity. The results of this  
paper provided evidence for this idea and indeed have already played a 
role in some of the just mentioned papers. 
\footnote{$^\dagger$}{The  main results of this paper are contained
in the Ph.D thesis of reference [14].} It also encourages the conjecture 
of reference [25] namely that all these theories ${\cal V}_{\cal G}$ 
might be contained as special limits of a single theory.

In this paper we constructed the respective theories as non-linear  
realisations and then deduce the underlying Kac-Moody algebra that  
could be a symmetry of the extended theory. However, given the  
Kac-Moody algebra one can reverse the process and construct its  
non-linear realisation at low levels. This is essentially the same  
calculation,  but read in reverse. Indeed, one can check that the  
generators in $D_8^{+++}$ at low levels, which were given in reference  
[25], are exactly those required to give rise to the fields of $N=1$  
supergravity and that the corresponding group element is that of  
equation (1.5).

\medskip
{\bf {Acknowledgments}}
\medskip
IS would like to thank Andr\'e Miemiec, who has helped to calculate
the equations of motion for various supergravity actions. This work is
partly supported by BSF - American-Israel Bi-National Science
Foundation, the Israel Academy of Sciences - Centers of Excellence
Program, the German-Israel Bi-National Science Foundation, and the
European RTN-network HPRN-CT-2000-00122.

\medskip
{\bf {References}}
\medskip
\parskip 0pt
\item{[1]} W. ~Nahm, {\it "Supersymmetries and their representations"},
     Nucl. Phys. {\bf B135} (1978), 149, E. Cremmer, B. Julia, J. Scherk,
     {\it Supergravity theory in 11 dimensions}, Phys. Lett. {\bf B76}
     (1978), 409

\item{[2]} I.C.G. Campbell, P. West, {\it N=2 d=10 nonchiral supergravity
     and its spontaneous compactifications}, Nucl. Phys. {\bf B243} (1984),
     112; M. Huq, M. Namanzie, {\it Kaluza-Klein supergravity in ten
     dimensions}, Class. Quant. Grav. {\bf 2} (1985); F. Giani, M. Pernici,
     {\it N=2 supergravity in ten dimensions}, Phys. Rev. {\bf D30} (1984),
     325

\item{[3]} J.H. Schwarz, P. West {\it Symmetries and Transformations of
     chiral N=2, D=10 supergravity}, Phys. Lett. {\bf B126} (1983), 301;
     J.H. Schwarz, {\it Covariant field equations of chiral N=2 D=10
     supergravity}, Nucl. Phys. {\bf B226} (1983), 269; P. Howe, P. West,
     {\it The complete N=2, d=10 supergravity}, Nucl. Phys. {\bf B238}
     (1984), 181

\item{[4]} L. Brink, J.H. Schwarz, J. Scherk, {\it Supersymmetric
     Yang-Mills theories}, Nucl. Phys. {\bf B121} (1977), 77; F. Gliozzi,
     J. Scherk, D.I. Olive, {\it Supersymmetry, supergravity theories and
     the chiral spinor model}, Nucl. Phys. {\bf B122} (1977), 253; 
     A.H. Chamseddine, {\it Interacting supergravity in ten dimensions: 
     the role of the six-index gauge field}, Phys. Rev. {\bf D24}(1981) 
     3065; E. Bergshoeff, M. de Roo, B. de Wit and P. van Nieuwenhuizen, 
     {\it Ten-dimensional Maxwell-Einstein supergravity, its currents, and 
     the issue of its auxiliary fields}, Nucl. Phys. {\bf B195} (1982) 97; 
     E. Bergshoeff, M. de Roo, B. de Wit, {\it Conformal supergravity in 
     ten dimensions}, Nucl.Phys. {\bf B217}(1983) 143; G. Chapline, N.S. 
     Manton, {\it Unification of Yang-Mills theory and supergravity in 
     ten dimensions}, Phys. Lett. {\bf 120B} (1983) 105 

\item{[5]} Luis Alvarez-Gaume, M.A. Vazquez-Mozo, {\it Topics in string 
    theory and quantum gravity}, {\tt hep-th/9212006} and references 
    therein

\item{[6]} D.J. Gross, J.A. Harvey, E.J. Martinec, R. Rohm, {\it The
     heterotic string}, Phys. Rev. Lett. {\bf 54} (1985), 502; D.J. Gross,
     J.A. Harvey, E.J. Martinec, R. Rohm, {\it Heterotic string theory.1.
     the free heterotic string}, Nucl. Phys. {\bf B256} (1985), 253; D.J.
     Gross, J.A. Harvey, E.J. Martinec, R. Rohm, {\it Heterotic string
     theory.2. the interacting heterotic string}, Nucl. Phys. {\bf B267}
     (1986), 75

\item{[7]} S. Ferrara, J. Scherk, B. Zumino, {\sl Algebraic properties 
    of extended supersymmetry}, Nucl. Phys. {\bf B 121} (1977) 393; 
    E. Cremmer, J. Scherk, S. Ferrara, {\sl $SU(4)$ invariant 
    supergravity theory}, Phys. Lett. {\bf B 74} (1978) 61

\item{[8]} E. Cremmer, B. Julia, {\sl The $N=8$ supergravity
    theory. I. The Lagrangian}, Phys. Lett. {\bf B 80} (1978) 48

\item{[9]} E. Cremmer, B. Julia, H. L\"u, C. Pope, {\it Dualisation of
    dualities. II Twisted self-duality of doubled fields and superdualities},
    Nucl. Phys. {\bf B535} (1998), 242, {\tt hep-th/9806106}

\item{[10]}  P. West, {\it Hidden superconformal symmetry in {M}
    theory},  JHEP {\bf 08} (2000) 007, {\tt hep-th/0005270}

\item{[11]}  P. West, {\it E11 and {M}-theory},
    Class. Quant. Grav. 18 (2001) 4443, {\tt hep-th/0104081}

\item{[12]} I. Schnakenburg, P. West, {\it Kac-Moody Symmetries of IIB 
    Supergravity}, Phys. Lett. {\bf B517} (2001) 421, {\tt hep-th/0107181}

\item{[13]} I. Schnakenburg, P. West, {\it Massive IIA supergravity
    as a non-linear realisation}, Phys. Lett. {\bf B540} (2002)
    137, {\tt hep-th/0204207}

\item{[14]} I. Schnakenburg, {\it Symmetries of Supergravity
    Theories and Quantum Field Theories}, PhD thesis, University of
    London, 2002

\item{[15]} M. Gaberdiel, D. Olive, P. West, {\it A class of Lorentzian
    Kac-Moody algebras}, Nucl. Phys. {\bf B645} (2002) 403,
    {\tt hep-th/0205068}

\item{[16]} N. Lambert, P. West, {\it Coset symmetries in dimensionally
    reduced bosonic string theory}, Nucl. Phys. {\bf B615} 117, {\tt
    hep-th/0107209}

\item{[17]} A. Barut, R. Raczka, {\it Theory of group Representations and
    Applications}, Polish Scientific Publishers, Warszawa 1980

\item{[18]} A. Casher, F. Englert, H. Nicolai and A. Taormina, {\it
    Consistent superstrings as solutions of the D=26 bosonic string theory}  
    Phys.Lett {\bf B162 } (1985) 121; F. Englert, H. Nicolai, A. 
    Schellekens, {\it Superstrings from $26$ dimensions}, Nucl.Phys.   
    {\bf B274} (1986) 315; F. Englert, L. Houart, A. Taormina, 
    {\it Brane fusion in the bosonic string and the emergence of fermionic 
    strings},  {\bf JHEP 0108} (2001) 013, {\tt hep-th/0106235};  
    A. Chattaraputi,  F. Englert, L. Houart, A. Taormina, {\it The bosonic 
    mother of fermionic D-branes }, {\bf JHEP 0209} (2002) 037,  
    {\tt hep-th/0207238}.

\item{[19]} P. West, {\it Very-extended $E_8$ and $A_8$ at low levels,
    gravity and supergravity}, Class. Quant. Grav. {\bf 20} (2003) 2393,
    {\tt hep-th/0212291}

\item{[20]} J. Maharana, J.H. Schwarz, {\it Noncompact symmetries in
     string theory}, Nucl. Phys. {\bf B390} (1993) 3 {\tt 9207016}; 
     J.H. Schwarz, A. Sen, {\it Duality Symmetries of 4d Heterotic 
     Strings}, Phys.Lett. {\bf B312} (1993), 105, {\tt hep-th/9305185}

\item{[21]} T. Damour, M. Henneaux, {\it E(10), BE(10) and arithmetical 
    chaos in superstring cosmology}, Phys. Rev. Lett. {\bf 86} (2001) 4749, 
    {\tt hep-th/0012172}; T. Damour, M. Henneaux, B. Julia, H. Nicolai,  
    {\it Hyperbolic Kac-Moody algebras and chaos in Kaluza-Klein models}, 
    Phys. Lett. {\bf B509} (2001) 323, {\tt hep-th/0103094}.

\item{[22]} S. Elitzur, A. Giveon, D. Kutasov, E. Rabinovici, {\it
    Algebraic aspects of matrix theory on $T^d$ }, {\tt hep-th/9707217}; 
    N. Obers, B. Pioline, {\it U-duality and  M-theory, an algebraic 
    approach}, {\tt hep-th/9812139}; T. Banks, W. Fischler, L. Motl, 
    {\it Dualities versus singularities}, {\bf JHEP 9901} (1999) 019, 
    {\tt hep-th/9811194}

\item{[23]} T. Banks, L. Motl, {\it On the hyperbolic structure of moduli 
    spaces with 16 supercharges}, {\bf JHEP 9905} (1999), 15, 
    {\tt hep-th/9904008}

\item{[24]} I. Schnakenburg, A. Miemiec, {\it $E_{11}$ and spheric vacuum
    solutions of eleven- and ten dimensional supergravity theories},
    {\tt hep-th/0312096}

\item{[25]} A. Kleinschmidt, I. Schnakenburg, P. West, {\it Very extended
    Kac-Moody algebras and their interpretation at low levels},
    {\tt hep-th/0309198}

\item{[26]} F. Englert, L. Houart, A. Taormina, P. West, {\it The
    symmetries of M-theory}, {\bf JHEP 0309} (2003), 20, 
    {\tt hep-th/0304206}; F. Englert, L. Houart, P. West, {\it 
    Intersection rules, dynamics and symmetries}, {\bf JHEP 0308} (2003) 
    025, {\tt hep-th/0307024}; F. Englert, L. Houart, {\it 
    ${\cal G}^{+++}$ invariant formulation of gravity and M-theories: 
    Exact BPS solutions}, {\tt hep-th/0311255}


\end